\documentclass[aps,pra,letterpaper,10pt,twocolumn,showpacs,superscriptaddress,notitlepage]{revtex4-2}

\usepackage[utf8]{inputenc}
\usepackage{amsfonts,amssymb,amsmath,amsthm}
\usepackage{url}
\usepackage{dsfont}
\usepackage{bbm}
\usepackage[dvipsnames, table]{xcolor}
\usepackage{nicefrac}
\usepackage{setspace}
\usepackage{tabularx,booktabs}
\usepackage{makecell}
\usepackage{comment}
\usepackage[normalem]{ulem}
\usepackage{siunitx}
\usepackage{physics}
\usepackage{comment}
\usepackage{graphicx}
\usepackage{multirow}
\usepackage[colorlinks=true, citecolor=blue,urlcolor=blue,linkcolor=blue,filecolor=black]{hyperref}
\renewcommand{\selectlanguage}[1]{}
\usepackage{siunitx}

\usepackage{ragged2e}

\makeatletter
\renewcommand{\fnum@figure}{\justifying\figurename~\thefigure}
\renewcommand{\fnum@table}{\justifying\tablename~\thetable}

\makeatother

\DeclareSIUnit{\pulse}{pulse}
\DeclareSIUnit{\sample}{Sa}
\DeclareSIUnit{\snu}{SNU}
\DeclareSIUnit{\symbol}{symbol}
\DeclareSIUnit{\baud}{Baud}

\usepackage{todonotes}

\definecolor{andreacol}{HTML}{0e9616}

\definecolor{giuliocol}{HTML}{00BFFF}

\definecolor{giuliocolbkg}{HTML}{00BFFF}

\definecolor{tomcol}{HTML}{ff00ff}
\definecolor{tomcolbkg}{HTML}{ffa5ff}

\usepackage{glossaries-extra}
\setabbreviationstyle[acronym]{long-short}
\glsdisablehyper
\newacronym{qrng}{QRNG}{Quantum Random Number Generator}
\newacronym{qkd}{QKD}{Quantum Key Distribution}
\newacronym{dv}{DV}{Discrete Variables}
\newacronym{cv}{CV}{Continuous Variables}
\newacronym{pic}{PIC}{Photonic Integrated Circuit}
\newacronym{plc}{PLC}{Planar Lightwave Circuit}
\newacronym{soi}{SOI}{Silicon-On-Insulator}
\newacronym{inp}{InP}{Indium Phosphide}
\newacronym{cmos}{CMOS}{Complementary Metal-Oxide-Semiconductor}
\newacronym{flm}{FLM}{Femtosecond Laser Micromachining}
\newacronym{cots}{COTS}{Commercial-Off-The-Shelf}
\newacronym{mzi}{MZI}{Mach-Zehnder Interferometer}
\newacronym{dc}{DC}{Directional Coupler}
\newacronym{tops}{TOPS}{Thermo-Optical Phase Shifter}
\newacronym{lo}{LO}{Local Oscillator}
\newacronym{llo}{LLO}{Local-Local Oscillator}
\newacronym{bpd}{BPD}{Balanced Photo-Detector}
\newacronym{ecl}{ECL}{External Cavity Laser}
\newacronym{pbs}{PBS}{Polarizing Beam Splitter}
\newacronym{pc}{PC}{Polarization Controller}
\newacronym{fpga}{FPGA}{Field-Programmable Gate Array}
\newacronym{qpsk}{QPSK}{Quadrature Phase-Shift-Keying}
\newacronym{psk}{PSK}{Phase-Shift-Keying}
\newacronym{qam}{QAM}{Quadrature Amplitude Modulation}
\newacronym{awg}{AWG}{Arbitrary Waveform Generator}
\newacronym{hpf}{HPF}{High-Pass Filter}
\newacronym{voa}{VOA}{Variable Optical Attenuator}
\newacronym{bs}{BS}{Beam Splitter}
\newacronym{pm}{PM}{Power Meter}
\newacronym{cmrr}{CMRR}{Common-Mode Rejection Ratio}
\newacronym{fwhm}{FWHM}{Full Width at Half-Maximum}
\newacronym{povm}{POVM}{Positive Operator-Valued Measurement}
\newacronym{skr}{SKR}{Secret Key Rate}
\newacronym{rrc}{RRC}{Root-Raised-Cosine}
\newacronym{pcs}{PCS}{Probabilistic Constellation Shaping}
\newacronym{tia}{TIA}{Trans-Impedance Amplifier}
\newacronym{vbs}{vBS}{Variable Beam Splitter}
\newacronym{psd}{PSD}{Power Spectral Density}
\newacronym{dsp}{DSP}{Digital Signal Processing}

\begin{document}

\title{{High-Performance Heterodyne Receiver for Quantum Information Processing \\ in a Laser Written Integrated Photonic Platform}}

\author{Andrea Peri}
\thanks{These two authors contributed equally}
\affiliation{Dipartimento di Ingegneria dell’Informazione, Università degli Studi di Padova, Via Gradenigo 6B, 35131 Padova, Italy}

\author{Giulio Gualandi}
\thanks{These two authors contributed equally}
\affiliation{Dipartimento di Fisica, Politecnico di Milano, Piazza Leonardo da Vinci 32, 20133 Milano (Italy)}
\affiliation{Istituto di Fotonica e Nanotecnologie, Consiglio Nazionale delle Ricerche (CNR), Piazza Leonardo da Vinci 32, 20133 Milano (Italy)}

\author{Tommaso Bertapelle}
\affiliation{Dipartimento di Ingegneria dell’Informazione, Università degli Studi di Padova, Via Gradenigo 6B, 35131 Padova, Italy}

\author{Mattia Sabatini}
\affiliation{Dipartimento di Ingegneria dell’Informazione, Università degli Studi di Padova, Via Gradenigo 6B, 35131 Padova, Italy}

\author{Giacomo Corrielli}
\affiliation{Istituto di Fotonica e Nanotecnologie, Consiglio Nazionale delle Ricerche (CNR), Piazza Leonardo da Vinci 32, 20133 Milano (Italy)}

\author{Yoann Piétri}
\affiliation{Dipartimento di Ingegneria dell’Informazione, Università degli Studi di Padova, Via Gradenigo 6B, 35131 Padova, Italy}

\author{Davide Giacomo Marangon}
\affiliation{Dipartimento di Ingegneria dell’Informazione, Università degli Studi di Padova, Via Gradenigo 6B, 35131 Padova, Italy}
\affiliation{Padua Quantum Technologies Research Center, Universit\`a degli Studi di Padova, via Gradenigo 6B, IT-35131 Padova, Italy}

\author{Giuseppe Vallone}
\affiliation{Dipartimento di Ingegneria dell’Informazione, Università degli Studi di Padova, Via Gradenigo 6B, 35131 Padova, Italy}
\affiliation{Istituto di Fotonica e Nanotecnologie - CNR, Via Trasea 7, 35131 Padova, Italy}
\affiliation{Padua Quantum Technologies Research Center, Universit\`a degli Studi di Padova, via Gradenigo 6B, IT-35131 Padova, Italy}

\author{Paolo Villoresi}
\affiliation{Dipartimento di Ingegneria dell’Informazione, Università degli Studi di Padova, Via Gradenigo 6B, 35131 Padova, Italy}
\affiliation{Istituto di Fotonica e Nanotecnologie - CNR, Via Trasea 7, 35131 Padova, Italy}
\affiliation{Padua Quantum Technologies Research Center, Universit\`a degli Studi di Padova, via Gradenigo 6B, IT-35131 Padova, Italy}

\author{Roberto Osellame}
\affiliation{Istituto di Fotonica e Nanotecnologie, Consiglio Nazionale delle Ricerche (CNR), Piazza Leonardo da Vinci 32, 20133 Milano (Italy)}

\author{Marco Avesani}
\email[\\Corresponding email: ]{marco.avesani@unipd.it}
\affiliation{Dipartimento di Ingegneria dell’Informazione, Università degli Studi di Padova, Via Gradenigo 6B, 35131 Padova, Italy}
\affiliation{Padua Quantum Technologies Research Center, Universit\`a degli Studi di Padova, via Gradenigo 6B, IT-35131 Padova, Italy}

\begin{abstract}
Continuous-Variable Quantum Key Distribution (CV-QKD) and Quantum Random Number Generation (CV-QRNG) are critical technologies for secure communication and high-speed randomness generation, exploiting shot-noise-limited coherent detection for their operation. 
Integrated photonic solutions are key to advancing these protocols, as they enable compact, scalable, and efficient system implementations.
In this work, we introduce Femtosecond Laser Micromachining (FLM) on borosilicate glass as a novel platform for producing Photonic Integrated Circuits (PICs) realizing coherent detection suitable for quantum information processing.
{Employing off-chip detectors, we} exploit the specific features of FLM to produce a PIC designed for CV-QKD and CV-QRNG applications.
The PIC features fully adjustable optical components that achieve precise calibration and reliable operation under protocol-defined conditions.
The device exhibits low insertion losses ($\leq 1.28$ dB), polarization-insensitive operation, and a Common-Mode Rejection Ratio (CMRR) exceeding 73 dB. These characteristics allowed the experimental realization of a source-device-independent CV-QRNG with a secure generation rate of 42.74 Gbps and a \gls{qpsk}-based CV-QKD system achieving a secret key rate of 3.2 Mbit/s.
Our results highlight the potential of FLM technology as an integrated photonic platform, paving the way for scalable and high-performing quantum communication systems.
\end{abstract}

\maketitle

\section{Introduction}
In recent years, quantum information processing has witnessed exponential advancements, especially in the fields of \glspl{qrng} and \gls{qkd}.
The former aims to provide true randomness, whereas the latter enables
information-theoretic secure communications between parties over an untrusted channel.
\gls{qrng} and \gls{qkd} protocols can be classified as either \gls{dv} or \gls{cv}, based on the degrees of freedom of the underlying quantum systems.
The latter approach, based on coherent detection, is particularly attractive in terms of performance and compatibility with the telecommunication industry.
In fact, \gls{cv}-\glspl{qrng} can achieve generation rates of several \si{Gbps} \cite{marangon2017source, Avesani2018, gehring2021homodyne}, and \gls{cv}-\gls{qkd} can reach
higher key rates compared to \gls{dv} systems, although for shorter distances~\cite{zhang2024continuous, Wang2024, hajomer2025}.

Along with the progress made in the aforementioned areas, there have also been significant improvements in the methods and materials used for their implementation.
Coherent detection systems consist of an optical front end connected to an electrical back end, which together enable the realization of either balanced homodyne or heterodyne detection.
Research has focused on miniaturizing these systems by leveraging integrated photonics to develop high-performance, scalable and mass-manufacturable \glspl{pic}~\cite{wang2020integrated,moody20222022,aldama2022integrated,luo2023recent}.
The level of integration that \glspl{pic} can achieve depends on the substrate used.
For example, \glspl{plc} realized in silica cannot monolithically incorporate either active components or pin photodiodes, which must be externally coupled~\cite{huang2019integrated}.
Silicon photonics instead relies on \gls{soi} wafers as its semiconductor substrate, making \gls{soi} technology compliant with \gls{cmos} fabrication techniques.
\gls{soi} chips can include complex waveguide structures and active control elements, such as Mach-Zehnder, multimode interferometers and phase shifters.
Light sources can only be externally coupled, but pin photodiodes can be hybridly (InGaAs)~\cite{bai202118, Bertapelle:25} or monolithically (Ge) integrated~\cite{raffaelli2018homodyne, zhang2019integrated, jia2023silicon, pietri2024experimental, li2024chip}.
Moreover, the compatibility of \gls{soi} \glspl{pic} with standard \gls{cmos} processes enables the coexistence of silicon electronics and photonics on the same substrate.
In fact, custom transimpedance amplifiers were recently integrated into silicon chip~\cite{tasker2024bi}, a solution that may enhance the capabilities of already existing high-performance \gls{cv}-\gls{qkd} receivers operating at tens of \si{\giga\hertz}~\cite{hajomer2024continuous, hajomer2025} and \gls{cv}-\glspl{qrng} achieving rates of hundreds of \si{Gbps}~\cite{bruynsteen2023100, bomhals202364}.
Another platform recently developed is Indium Phosphide (InP), which supports the monolithic integration of lasers and phodiodes alongside all the active and passive optical elements common to \gls{soi} technology.
This makes InP capable of reaching the highest degree of integration of optical components~\cite{kincaid2023source, Aldama2025}.
However, both \gls{soi} and InP platforms typically exhibit significant polarization dependence, requiring careful polarization control and limiting their applicability in scenarios where polarization is not fixed or cannot be easily managed.
In addition, miniaturization is not the only figure of merit when developing integrated photonics for quantum technologies.
Indeed, one of the key challenges in quantum communication protocols is the minimization of optical losses, especially at the receiver side
for \gls{qkd}, because they result in irreversible information loss.

In this work, we introduce \gls{flm} on borosilicate glass as a powerful platform to manufacture \glspl{pic} designed for high-performance quantum coherent detection.
With \gls{flm}, waveguides can be directly written within transparent materials with three-dimensional features and essentially at any wavelength across the VIS–NIR range by focusing short and intense laser pulses for direct core writing ~\cite{Corrielli2021}.
Although offering lower miniaturization capabilities compared to other photonic platforms, due to the relatively low refractive index contrast achievable (up to $10^{-2}$), \gls{flm} provides enhanced modularity, low losses, and cost-effective fabrication. The close match between waveguide modes (a few $\si{\micro\meter\squared}$) and standard optical fibers ensures efficient edge coupling, with interface losses as small as $0.2$~dB and propagation losses around $0.1$~dB/cm at $1550$~nm. These features make glass-based devices highly suitable for quantum photonic experiments, where minimizing attenuation is essential.

Furthermore, the technology enables the realization of polarization-insensitive devices: any polarization state injected in the chip undergoes the same unitary transformation, since directional couplers implement the same splitting ratio for any polarization state and the weak residual birefringence ($10^{-6}$–$10^{-5}$, mainly due to core-shape anisotropy) induces the same rotation across all paths. This intrinsic polarization insensitivity allows modular interconnection of multiple devices without the need for precise polarization control, in contrast to other integrated platforms. These features, which have already been exploited in various quantum information processing tasks~\cite{Marshall:09, Crespi_AL2013, schell2013three, crespi2015particle, seri2018laser, chen2019laser, Hoch_2022, Sax2023}, not only simplify system design but also unlock new possibilities beyond the reach of conventional photolithographic techniques.
Here we demonstrate the suitability and versatility of glass \gls{flm} for the implementation of quantum coherent detection for state-of-the-art \gls{cv}-\gls{qrng} and \gls{cv}-\gls{qkd} protocols.
Specifically, we developed a \gls{pic} that serves as a low-loss polarization-insensitive optical front-end of a heterodyne receiver. 
Such properties allow for efficient coupling of the chip with \gls{cots} fiber-based lasers and balanced detectors.
Moreover, the \gls{pic} was designed with tunable capabilities: the relative phase of the heterodyne measurements, as well as their balancing, can be fully adjusted.
This not only enhances the performance of its electronic back-end, but also allows the system to operate under conditions that best align with the theoretical requirements of CV protocols, thereby ensuring maximum reliability and assurance in their implementation.

In the next sections, we will detail the design of the \gls{pic} and its characterization, Section \ref{Sec:DFC}; the implementation of a \textsl{source-device-independent} random number generation protocol, Section \ref{Sec:QRNG}; the implementation of the \gls{qpsk} \gls{cv}-\gls{qkd} protocol, Section \ref{Sec:QKD}.
Then, in Section \ref{Sec:Conclusion} we will discuss our findings.

\section{PIC Fabrication, Design and Characterization}\label{Sec:DFC}

\subsection{PIC fabrication and design}

The glass-based optical circuit was manufactured using the \gls{flm} microfabrication technique~\cite{Corrielli2021}.
This process tightly focuses ultra-short laser pulses inside a glass substrate, triggering non-linear absorption that induces a permanent and localized increase in its refractive index,  primarily driven by ion migration during the transient melting and rapid resolidification caused by the laser irradiation \cite{fernandez2018bespoke}.
Such a process is confined to the pulse focal volume, allowing for the direct writing of waveguides upon translation of the substrate relative to the laser focus.
Moreover, the possibility to move the sample in three dimensions enables the realization of out-of-plane photonic circuits.
This versatile technique is suitable for fast prototyping and grants full control over multiple parameters during fabrication.
Indeed, by changing the irradiation parameters, it is possible to optimize the properties of waveguides, such as losses and birefringence, for any desired wavelength.
Furthermore, since the waveguide geometry is controlled with an accuracy of tens of nanometers, directional couplers can be fabricated with splitting ratio tolerances $<\SI{1}{\percent}$.
This enables the realization of high-quality balanced directional couplers and, consequently, \glspl{mzi} with high extinction ratios.
The latter is a crucial factor that directly impacts the device performance, as will be discussed in Sections \ref{Sec:QRNG} and \ref{Sec:QKD}.

\begin{figure*}[t!]
    \centering
    \includegraphics[width=1.\linewidth]{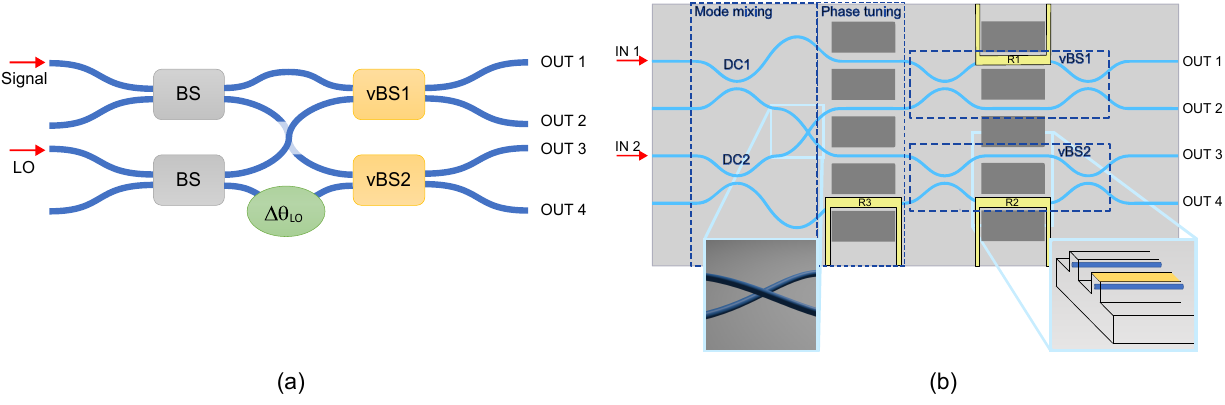}
    \caption{\textbf{FLM-written PIC layout}.
    (a) Conceptual layout of the \textit{tunable optical hybrid}, where we can independently control the phase shift $\Delta\theta_{\rm LO}$ and the splitting ratio of the variable beam splitters (vBSs). The two red arrows indicate the inputs for the signal and \glsxtrlong{lo}, which are then processed in balanced beam splitters (BSs), waveguide crossing, and variable beam splitters (vBSs).
    When the phase $\Delta\theta_{\rm LO}$ is $\pi/2$ and the vBSs are perfectly balanced, the device acts as a \textit{$\SI{90}{\degree}$ optical hybrid}. 
    (b) Schematic layout of the FLM-written photonic device. The dashed rectangle on the left highlights the mode mixing, where the two central modes swap by crossing one over the other (see left inset) thanks to the three-dimensional capabilities of \gls{flm}.
    The central dashed rectangle evidences the area where the thermal shifter tunes the $\Delta\theta_{\rm LO}$ phase shift. The two dashed rectangles on the right mark the locations of the vBSs, implemented through reconfigurable Mach-Zehnder interferometers, used to balance the device's outputs. The thermo-optic phase shifters (associated to resistances R1, R2, and R3) are depicted in yellow, and the dark gray rectangles represent the trenches fabricated by \gls{flm} alongside the waveguides to provide increased thermal insulation (see right inset), thereby reducing power dissipation per micro-heater and minimizing thermal cross-talk.
   }
    \label{fig: Layout_chip}
\end{figure*}

\begin{table}[ht]
    \centering
    \begin{tabular}{cccc}
        \toprule
        \multicolumn{4}{c}{\textbf{DEVICE SPECIFICATIONS}} \\
        \midrule
        \textbf{Substrate} & \multicolumn{3}{c}{Borosilicate Glass Eagle XG} \\
        \multirow{2}{*}{\textbf{Insertion Losses}} & IN1 & \SI{1.05}{\dB} & \\
        & IN2 & \SI{1.28}{\dB} & \\
        \multirow{3}{*}{\textbf{DC Reflectivities}} & Polarization & DC1 & DC2 \\
        & H & 0.502 & 0.506 \\
        & V & 0.502 & 0.502 \\
        \multirow{2}{*}{\textbf{TOPS Power Dissipation}} & & $2\pi$ & \\
        & & \SI{70}{\milli\watt} & \\
        \bottomrule
    \end{tabular}
    \caption{\textbf{FLM-written PIC specifications}.
    The table presents the insertion losses for inputs 1 and 2, as indicated in Fig.~\ref{fig: Layout_chip}(b), along with the splitting ratio for the directional couplers relative to IN1 (DC1) and IN2 (DC2), for both horizontal and vertical polarization.
    Finally, the power dissipation required to induce a $2\pi$ phase shift is reported.}
    \label{tab:device_specifications}
\end{table}

The device presented in this work was fabricated using femtosecond laser pulses with a wavelength of \SI{1030}{\nano\meter}, a repetition rate of \SI{1}{\mega\hertz}, and a pulse duration of \SI{180}{\femto\second}.
Laser light was focused into the volume of an Eagle XG borosilicate glass substrate through a water immersion objective with a numerical aperture of $0.5$, at a depth of $\SI{50}{\micro\meter}$ from the surface.
For the fabrication of single-mode waveguides, we adopted an energy of $\SI{500}{\nano\joule\per\pulse}$ and repeated four overlapping writing scans per waveguide, each at a translation speed of $\SI{7}{\milli\meter/\second}$.
To enhance the optical quality of the waveguides, we performed a thermal annealing step to relax any stresses formed during the fabrication process.
This treatment improves the optical mode confinement, allowing bending radii of \SI{10}{\milli\meter}~\cite{Arriola2013}.

As illustrated in Fig.~\ref{fig: Layout_chip}(a), the PIC implements a {\it tunable optical hybrid} composed of two 50:50 \glspl{bs}, 
a variable phase shifter $\Delta \theta_{\rm LO}$, and two \glspl{vbs} implemented as Mach-Zehnder interferometers.
When $\Delta \theta_{\rm LO}=\pi/2$ and the \glspl{vbs} are balanced, the device corresponds to a $90^\circ$ optical hybrid, a key component in quantum and classical coherent detection.
An image illustrating the \gls{pic}'s layout is reported in Fig.~\ref{fig: Layout_chip}(b).
The chip has a footprint of $1.5 \times \SI{3.7}{\centi\meter}$ with its main specifications listed in Table~(\ref{tab:device_specifications}).
The insertion losses of the fiber-pigtailed device relative to IN1 and IN2 were measured to be 1.05 and 1.28 dB, with negligible difference ($< 1\%$) with respect to input polarization. These values account for propagation and bending loss of the waveguides, coupling losses due to mode mismatch between fiber and waveguide fundamental modes and fiber-to-fiber mating at the connector.
The geometry of all \glspl{dc} was optimized to achieve a splitting ratio close to 50:50.
For each \gls{dc}, we also reported in Table~(\ref{tab:device_specifications}) the measurement of their reflectivity for both horizontal (H) and vertical (V) polarization.
From such values, we can infer that they are almost polarization insensitive, deviating from ideality by $<\SI{1}{\percent}$.
To realize the tunable optical hybrid, we exploited the three-dimensional capabilities of \gls{flm}.
\begin{figure*}[t!]
    \centering
    \includegraphics[width=\textwidth]{images/Figure02.pdf}
    \caption{\textbf{Experimental setup}. The figure details the optical components used for both the \gls{pic} characterization and the implementation of the \gls{cv} quantum protocols.
    The optical source provides a coherent beam with fixed power and polarization.
    Such optical power is split and undergoes different paths, generating the \gls{lo} and the phase-modulated quantum signal $\ket{\alpha}$.
    These modes are coupled at the \gls{pic}'s inputs and then converted into electrical signals by two \glsxtrlongpl{bpd} to measure the quadratures of the quantum signal.
    The configuration of the \gls{pic} as an optical heterodyne receiver consists of balancing the chip's outputs and setting the tunable optical hybrid as a \SI{90}{\degree} one.
    While the balancing procedure is performed just with the \gls{lo} optical beam, setting the hybrid required a phase-modulated classical signal along with the \gls{lo}.
    Once configured, the device is exploited as the optical receiver for \gls{cv}-\gls{qrng} and \gls{cv}-\gls{qkd} protocols.
    \glsxtrshort{awg}: \glsxtrlong{awg}, ISO: optical isolator, \glsxtrshort{pc}: \glsxtrlong{pc}, \glsxtrshort{pbs}: \glsxtrlong{pbs}, \glsxtrshort{bs}: \glsxtrlong{bs}, \glsxtrshort{pm}: \glsxtrlong{pm}, \glsxtrshort{fpga}: \glsxtrlong{fpga}, AMP: RF amplifier, \glsxtrshort{voa}: \glsxtrlong{voa}, ATT: optical attenuator, \glsxtrshort{bpd}: \glsxtrlong{bpd}.}
    \label{fig: optical setup}
\end{figure*}
As shown in Fig.~\ref{fig: Layout_chip}(b), the two central waveguides cross each other at different depths, with $\SI{15}{\micro\meter}$ vertical spacing, resulting in no crosstalk between them.
Control over the optical phases, required for the proper operation of the PIC as a $\SI{90}{\degree}$ hybrid and for tuning the splitting ratio of the \glspl{vbs}, is achieved through integrated \glspl{tops}.
Each \gls{tops} is fabricated by ablating with the femtosecond laser a $\SI{100}{\nano\meter}$-thick Cr+Au layer, deposited on the substrate surface, to pattern its electrical pads and micro-heaters~\cite{Pont2024}.
In addition, as shown in Fig. \ref{fig: Layout_chip}, insulating trenches were ablated aside the waveguides to reduce thermal crosstalk and power consumption.
In particular, these \glspl{tops} introduce a $2\pi$ phase shift using $\SI{70}{\milli\watt}$ of dissipated electrical power.
The TOPSs are addressed by electrically connecting the gold contact pads to a custom-made PCB using conductive epoxy, and driven via a voltage source connected to the PCB.
More details on the fabrication can be found in~\cite{Albiero2022, Ceccarelli2020}.

The \gls{flm}-written \gls{pic} is designed to implement a phase-diverse heterodyne coherent detection.
Therefore, it must be ensured that the quantum signal properly interferes with a strong coherent beam (strong enough that its quantum behavior can be neglected and thus described classically), commonly referred to as the Local Oscillator (LO).
This is accomplished through the chip's optical routing and correct tuning of its TOPS voltages.
In our case, the quantum signal and \gls{lo}, entering the \gls{pic} through IN1 and IN2 respectively, are split by two 50:50 \glspl{dc}.
Next, each quantum signal path interferes with one LO branch via the vBSs, tuned as balanced beam splitters.
The latter is obtained by inducing the correct phase shift with \gls{tops} R1 (vBS1) and R2 (vBS2).
The optical signals from each vBS are connected off-chip to a pair of \glspl{bpd}, for the simultaneous detection of two quadratures of the input quantum state.
Moreover, to ensure orthogonality of the detected quadratures, one of the aforementioned \gls{lo} branches is phase-shifted on-chip by $\pi/2$ with \gls{tops} R3.
This design ensures flexibility and accurate control over the \gls{pic} operation, boosting the heterodyne receiver's performance.

\subsection{Measurement setup}
\label{sec: experimental_setup}

\begin{figure*}[t!]
    \centering
    \includegraphics[width=1.\linewidth]{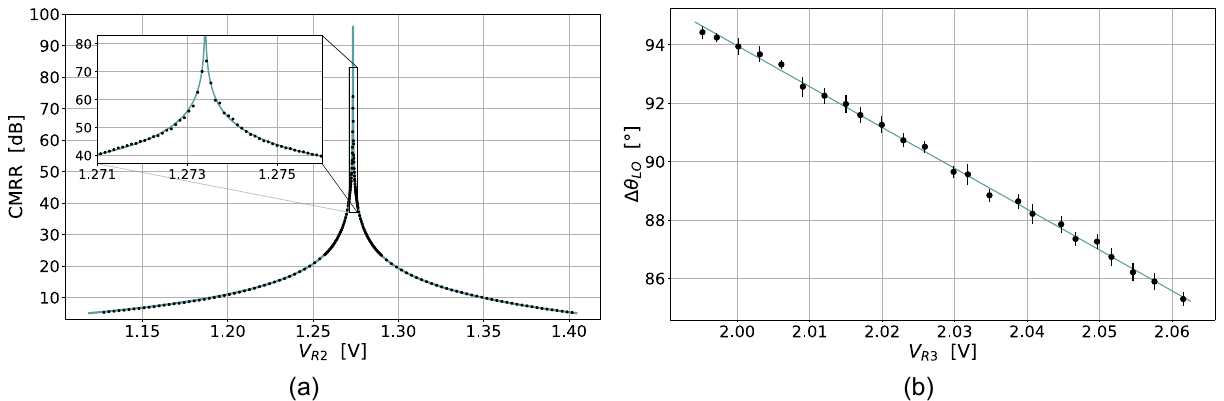}
    \caption{\textbf{Characterization of the \gls{pic} optical configuration}. (a) \gls{cmrr} response of the variable beam-splitter vBS2 with respect to the voltage applied to R2 \gls{tops}.
    The experimental data closely match the theoretical model described in Eq.~(\ref{eq: CMRR_model}).
    By fine-tuning $V_{R2}$, the device achieved a \gls{cmrr} as high as $\sim\SI{73.8}{\dB}$.
    The vBS1 tunable \gls{bs} displays a similar trend and performance.
    (b) Phase shift applied to the \gls{lo} by \gls{tops} R3 as a function of the driving voltage $V_{\rm R3}$.
    In the region of $\Delta\theta_{\rm LO}\approx \SI{90}{\degree}$, we estimate from the fitted linear coefficient a phase-voltage sensitivity of \SI{0.14}{\degree\per\milli\volt}, quantifying the TOPS response to the applied voltage.}
    \label{fig: het_characterization}
\end{figure*}

In Fig.~\ref{fig: optical setup}, we reported the schematic of the experimental setup used to operate the heterodyne optical receiver.
The latter was designed to allow the characterization of the \gls{pic} and the implementation of both the \gls{qrng} and \gls{qkd} experiments, avoiding unnecessary hardware reconfigurations.
A $\SI{1550}{\nano\meter}$ \gls{ecl} (Thorlabs SFL1550S) with a linewidth of $\sim\SI{50}{\kilo\hertz}$ was used to generate the optical signal and the \glsxtrlong{lo}.
We cascaded the laser with an optical isolator (ISO) and a \gls{pbs}, whose optical axes do not require to be aligned with respect to the \gls{pic} given its polarization transparency.
The former component prevents back-scattered light from causing stability issues with the \gls{ecl}, while the latter avoids possible polarization drifts.
To ensure all light is coupled to the \gls{pbs} branch leading to the optical heterodyne, we used a \gls{pc} before the \gls{pbs} and monitored the optical power of its other arm with a power-meter (PM-1).
The \gls{pc} is adjusted to minimize the power reading on PM-1.
Following the \gls{pbs}, a 99:1 Beam Splitter (BS-1) directs \SI{1}{\percent} and \SI{99}{\percent} of the laser light to the \gls{pic}'s signal and \gls{lo} port, respectively.
The light traveling through the \SI{1}{\percent}-arm (Signal line) is modulated using a polarization-insensitive phase encoder~\cite{Sabatini2025} incorporating a lithium niobate phase modulator (iXblue MPZ-LN-10), driven by a \gls{fpga} (AMD Xilinx RFSoC 4×2) working as an arbitrary waveform generator.
The encoder has two different purposes throughout this work.
On one hand, it is used to generate the classical signals required to configure the chip as a $90^\circ$ optical hybrid.
On the other hand, in the CV-QKD experiment, it will serve as part of the optical transmitter.
After the phase encoder, a polarization controller (PC-2) maximizes the interference between the modulated signals and the receiver LO at the PIC input.
A 99:1 beam splitter (BS-2) then splits the optical power: the $1\%$ branch is routed to the heterodyne signal port through a short fiber and a $1.7~\si{dB}$ optical attenuator, while the $99\%$ branch is directed to the powermeter \mbox{(PM-2)} used to monitor the signal power.
The LO line (BS-1 $99\%$-arm) includes an optical delay line ($\Delta L$) to match its overall length to that of the Signal line, thereby minimizing phase instabilities at the receiver end.
It also features an electrically controlled variable optical attenuator (VOA-1) to adjust the LO power entering the PIC.
To estimate the latter quantity, a 99:1 beam splitter (BS-3) redirects $1\%$ of the \gls{lo} light to the power-meter (PM-3).

The phase-modulated signal and \gls{lo} constitute the \gls{pic}'s inputs, where they interfere according to the \glspl{tops} configuration.
The latter is set by a multi-channel voltage generator with sub-mV resolution.
In the context of \gls{qrng} and \gls{cv}-\gls{qkd} operation, the \glspl{tops} are tuned to introduce the correct phases required for the heterodyne architecture, which samples two orthogonal quadratures of the quantum signal.
Thermal control of the \gls{pic}, achieved via a Peltier element, ensures robust operation of the heterodyne configuration.
Finally, the optical signals at the \gls{pic}'s outputs are processed into electrical signals by a pair of \glspl{bpd} with \SI{2.5}{\giga\hertz} bandwidth (Thorlabs PDB780CAC), filtered by an analog \SI{1}{\mega\hertz} \gls{hpf} to reject low-frequency noise, digitized by a $\SI{25}{\giga\sample\per\second}$ oscilloscope (Tektronix DPO70404C) with $8$-bit resolution per channel, and transferred to a computer for digital processing.

\subsection{PIC characterization}\label{ssec:pic_characterization}
Throughout this work, the \gls{pic} was employed as a heterodyne receiver to detect quantum signals at the shot-noise limit using \glspl{bpd}.
This requires to appropriately configure the chip's \glspl{tops}.
In particular, the \gls{lo}'s phase entering vBS2 is set (through R3) to be $\pi/2$ out-of-phase, and the vBSs are finely balanced using R1 and R2.
This fine control of the optical components within the vBSs allows us not only to suppress the common-mode classical noise introduced by the laser but also to increase the \gls{lo} power at the heterodyne input, ultimately boosting the performance of the quantum devices.
These performance enhancements are possible by configuring each output of the system, composed of the integrated vBS and the off-chip BPD, to maintain a 50:50 splitting ratio.
This is accomplished by maximizing the system's \gls{cmrr}, which is estimated by directly modulating the \gls{ecl} laser with a $\SI{500}{\kilo\hertz}$ sine-wave electrical signal while blocking the Signal line.
As detailed in Appendix~\ref{App:pic_config}, the \gls{cmrr} is computed by evaluating the ratio of the \gls{bpd}'s electrical power at $\SI{500}{\kilo\hertz}$ when completely un-balancing and balancing the \glspl{vbs}.
\begin{figure*}[t!]
    \centering
    \includegraphics[width=1.\linewidth]{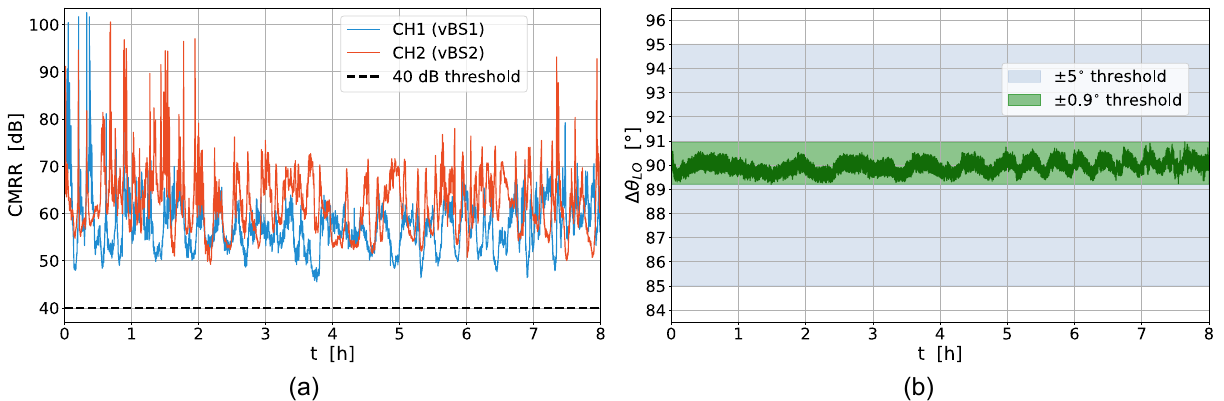}
    \caption{\textbf{Temporal stability of the \gls{pic} heterodyne configuration}. a) The \gls{cmrr} stability is shown for both the vBS.
    For quantum protocols implementations, a \gls{cmrr} lower threshold is set to $\SI{40}{\dB}$, since higher values do not lead to a significant improvement in receiver performance, as we will discuss in Section \ref{Sec:QRNG} and shown in Fig.~\ref{fig: min_entropy_calib_line}.
    Higher \gls{cmrr} values are consistently measured throughout the test duration.
    b) Stability of the \gls{lo} \SI{90}{\degree} phase shift. The plot compares the \gls{pic} phase stability (in green) with the $\pm \SI{5}{\degree}$ phase error commonly reported for commercial $\SI{90}{\degree}$ optical hybrid devices (in blue).
    Apart from temperature stabilization, no active mechanism was employed to maintain the device configuration.}
    \label{fig: PIC_stability}
\end{figure*}
Fig.~\ref{fig: het_characterization}(a) shows the relationship between vBS2 \gls{cmrr} and the driving voltage of the associated \gls{tops}.
As can be seen, it closely follows the trend:
\begin{equation}
    \text{CMRR}=-20\log\Big{(}\left|\sin\phi_\text{R2}\right|+\delta\Big{)}
    \label{eq: CMRR_model}\, .
\end{equation}
This equation corresponds to the standard sinusoidal response of a \gls{mzi} as a function of the interferometric phase $\phi_\text{R2}$, with the latter being polynomially dependent on the \gls{tops} voltage $V_\text{R2}$ \cite{ceccarelli2019}.
The parameter $\delta$ quantifies the maximum CMRR achievable by the system and it is estimated based on the system noise floor, as detailed in Appendix~\ref{App:pic_config}.

The experimental data exhibits a sharp peak located in the region where the vBS splitting ratio is close to the ideal 50:50, with a \gls{fwhm} of \SI{2.0}{\milli\volt} and a maximum measured \gls{cmrr} of $\sim\SI{73.8}{\dB}$.
The vBS1 shows an analogous behavior.
This indicates that a resolution of fractions of \si{\milli\volt} is required to drive the integrated \gls{tops} to ensure the \glspl{vbs} are highly balanced.
However, external factors such as small temperature and polarization drifts may affect the high \gls{cmrr} values reached while configuring the device, leading to a degradation of such a quantity over time.
As shown in Tab.~(\ref{tab:device_specifications}) 
the couplers in the PIC have an excellent polarization insensitivity; however, even a dependence on the third decimal place of the splitting ratio limits the \gls{cmrr} to approximately $\SI{30}{\dB}$.
Therefore, to enhance the stability of the \gls{cmrr} in the $\SI{70}{\dB}$ range, it is crucial to suppress polarization drifts.
For the aforementioned reasons, we thermally stabilized the \gls{pic} and placed a \gls{pbs} at the laser output to stabilize the polarization of the optical source.

To ensure that the heterodyne receiver measures two orthogonal quadratures, we acted on R3 to provide a phase shift $\Delta\theta_{\rm LO}$ of $\pi/2$ to the \gls{lo} entering vBS2.
To reach such a condition, by referring to Fig.~\ref{fig: optical setup}, the light in the Signal line was modulated with a triangular waveform covering the whole $2\pi$ range of the state's encoder.
In Fig.~\ref{fig: het_characterization}(b) we show the relation between $\Delta\theta_{\rm LO}$ and $V_{\rm R3}$ driving voltage.
Despite the polynomial dependence between the latter quantities mentioned above, a first-order approximation is sufficient to describe the behavior of the device in the $\SI{90}{\degree}$-neighborhood, demonstrating a phase-voltage sensitivity of $\SI{0.14}{\degree\per\milli\volt}$.

\begin{table*}
    \centering
    \scriptsize
    \begin{tabular}{ccc|ccccc}
        \toprule
         \multirow{2}{*}{Technology} & \multirow{2}{*}{Hybrid type} & \multicolumn{2}{c}{Hybrid losses} & \multirow{2}{*}{Detector efficiency} & \multirow{2}{*}{Overall efficiency} & \multirow{2}{*}{Phase deviation} & \multirow{2}{*}{Ref} \\
         & & On-chip & Coupling & & & & \\
         \midrule
         Micro-optics & \SI{90}{\degree} & \multicolumn{2}{c}{\SI{1}{\dB} (typ.)} & N/A & N/A & $\pm \SI{5}{\degree}$ & Exail COH90 \\
         Silicon & \SI{90}{\degree} & \multicolumn{2}{c}{\SI{3.19}{\dB}} & \SI{80}{\percent} (typ. Ge)$^\dagger$ & \SI{38.4}{\percent} & $\pm\SI{7.5}{\degree}$ &  Finisar CPRV\\
         \midrule
         Silicon & \SI{180}{\degree} & \multicolumn{2}{c}{\SI{5.1}{\dB}} & \SI{64}{\percent}$^\dagger$ & \SI{20}{\percent} & N/A & \cite{raffaelli2018homodyne}\\
         Silicon &\SI{180}{\degree} & \multicolumn{2}{c}{\SI{5}{\dB}}  & \SI{73}{\percent} (inferred)$^\dagger$ & \SI{49.8}{\percent} & N/A&  \cite{zhang2019integrated}\\
         Silicon & \SI{90}{\degree} &\multicolumn{2}{c}{\SI{2.5}{\dB}} & \SI{88}{\percent}$^\dagger$ & \SI{44}{\percent} & NR &  \cite{hajomer2024continuous}\\
         Silicon & \SI{180}{\degree} &\multicolumn{2}{c}{\SI{4}{\dB}} & \SI{64}{\percent}$^\dagger$ & \SI{22.7}{\percent} & N/A & \cite{bian2024continuousvariablequantumkeydistribution}\\
         GaInAsP/InP (*) &\SI{90}{\degree} & 0.5dB & 6.9dB & N/A & N/A & $\pm 5^\circ$& \cite{Jeong20101323} \\
         Si3N4 (*) &\SI{90}{\degree} & $<$ 1dB & NR & N/A & N/A & $<\SI{8}{\degree}$ & \cite{YU2020125620}\\
         Silicon &\SI{180}{\degree} & \SI{2.1}{\dB} & \SI{3}{\dB} & \SI{80}{\percent}$^\dagger$ & \SI{26}{\percent} & N/A & \cite{pietri2024experimental} \\
         FLM &\SI{90}{\degree} & \SI{0.65}{\dB} & \SI{0.4}{\dB} & $\SI{75}{\percent}$ (avg.) & \SI{59.3}{\percent} & $\pm\SI{0.9}{\degree}$ & This work \\
         \bottomrule
    \end{tabular}
    \caption{\textbf{Comparison of commercially available and \gls{pic}-based hybrids.} The two first entries correspond to commercially available hybrids. IL: Insertion Losses, including propagation losses. When the separation between insertion losses and coupling losses are not known, the overall hybrid losses are indicated.\\
    (*) The device was not used for quantum communication applications. \\
    ($^\dagger$) The photo-detectors are integrated on the chip. \\
    “NR”: Information not reported in the cited study.}
    \label{tab:comparison}
\end{table*}

Finally, we evaluated the device’s temporal stability over two eight-hour tests, monitoring the \glspl{vbs}' \gls{cmrr} and $\Delta\theta_{\rm LO}$.
The results are shown in Fig.~\ref{fig: PIC_stability}.
Fluctuations in the measured quantities are attributed to limitations of the PID control loop regulating the \gls{pic} temperature.
Nevertheless, despite these fluctuations, the \gls{cmrr} reads well above the \SI{40}{dB} threshold throughout the duration of the test.
As it will be discussed in Section~\ref{Sec:QRNG}, this level is more than sufficient to ensure reliable operation for both \gls{qrng} and \gls{cv}-\gls{qkd} applications.
Fig.~\ref{fig: PIC_stability}(b) shows a highly stable \SI{90}{\degree} optical hybrid operation, with a maximum phase fluctuation of $\pm \SI{0.9}{\degree}$ around a mean value of $\SI{90.0}{\degree}$.
The stability of the \gls{lo} phase is compared with the phase error generally reported for bulk, commercially-available \SI{90}{\degree} optical hybrid devices, which is in the order of $\pm \SI{5}{\degree}$~(Table~\ref{tab:comparison}).
These tests, along with the characterization reported above, guarantee an accurate and stable control over the PIC operation, showcasing the capabilities of our device.

Table~(\ref{tab:comparison}) compares the performance of our \gls{pic} with commercially-available devices and implementations reported in the literature based on \SI{90}{\degree} and \SI{180}{\degree} hybrids, mainly focusing on those used in \gls{qrng} and \gls{cv}-\gls{qkd} experiments.
This showcases the high input-to-output efficiency of our \gls{pic} compared to other platforms that suffer from higher coupling losses.
The phase deviation is also shown for \SI{90}{\degree} hybrids when the information was available.
Note that this value is usually given over the entire bandwidth, while in our case it corresponds to the value achieved at $\SI{1550}{\nano\meter}$.
However, the reconfigurability of our \gls{pic} allows to reach such precision also for different wavelengths.
In addition, our \gls{pic} could be combined with even higher-efficiency detectors.
For example, commercially available solutions with $\SI{1.2}{\ampere/\watt}$ of responsivity at $\SI{1550}{\nano\meter}$ would enable our implementation of a coherent receiver to reach an overall efficiency of \SI{77}{\percent}.
As can be seen on Table~(\ref{tab:comparison}), our \gls{flm} chip performs better than existing implementations in terms of optical losses and phase deviation.

\section{QRNG Configuration} \label{Sec:QRNG}
In the context of \glspl{qrng} for cryptographic applications, our \gls{pic} enables the implementation of heterodyne-based protocols by providing a stable coherent receiver with high \gls{cmrr} and precise control of the $\pi/2$ phase difference between the measured quadratures.
In this work, we implement the source-device independent protocol described in~\cite{Avesani2018}, which provides secure randomness regardless of the quantum state used, even if it is fully controlled by an attacker.
Indeed, even in the case of general attacks, the \gls{povm} structure of the optical heterodyne receiver lower-bounds the quantum conditional min-entropy:
\begin{equation}
    H_\text{min}\left( X \vert E \right) \geq -\log_2\left( \dfrac{\delta_x\delta_p}{\pi} \right)^n\, ,
\end{equation}
where $n$ is the number of rounds of the protocol and $\delta_{x,p}$ denote the device's phase-space resolution.
The quantity $H_\text{min}\left( X \vert E \right)$, in turn, upper-bounds the number of secure true random bits that can be extracted from a sequence of heterodyne measurement outcomes.
Additionally, compared to implementations belonging to the broader class of semi-device independent generators \cite{Ma2016}, our protocol of choice is considerably simpler to implement, it does not require external sources of randomness or real-time min-entropy analysis \cite{PhysRevA.104.062424} and can deliver significantly higher performances \cite{unbounded22}.

Before running the \gls{qrng}, the receiver must be characterized, specifically the device's phase-space resolution $\delta_{x,p}$ must be determined.
This characterization is essential to ensure the security of the generated numbers.
To this end, we followed the procedure outlined in the supplementary materials of~\cite{Avesani2018}.
Since the device's photo-detectors must operate within the linear regime, the quadrature measurement outcomes scale proportionally with the \glsxtrlong{lo}'s power $P_{LO}$.
Therefore, the phase-space resolutions $\delta_{x,p}$ of the receiver, expressed in shot-noise units, are given by:
\begin{equation}
    \delta_{x,p}= \dfrac{\delta_{VU}}{\sqrt{2 m_{x,p} P_{LO}+2q_{x,p}}} \leq \dfrac{\delta_{VU}}{\sqrt{2 m_{x,p} P_{LO}}}\, ,
\end{equation}
where $\delta_{VU}$ is the resolution in volt-units, $q_{x,p}$ accounts for the classical noise due to the receiver electronics, and $m_{x,p}$ is the proportionality coefficient relating the variance of the heterodyne signals to $P_{LO}$.
Notice that the latter upper-bound is obtained by un-trusting the classical electronic noise.
This effectively reduces the generator's rate but enhances the device security against the potential manipulation of such noise by an attacker.
Moreover, as the attacker's guessing probability decreases with increasing \gls{lo} power, it is desirable to maximize $P_{LO}$ while avoiding non-linear effects of the \glspl{bpd}.
As we demonstrate below, higher \gls{cmrr} values enable the use of stronger \gls{lo}, thus enhancing the generation rate.
After calibration, the \gls{qrng} performance was evaluated as a function of the \gls{cmrr}, and data was collected for randomness extraction and analysis with the standard NIST test suite \cite{nisttest}.

\subsection{Experimental setup}

\begin{figure*}[t!]
    \centering
    \includegraphics[width=1.0\linewidth]{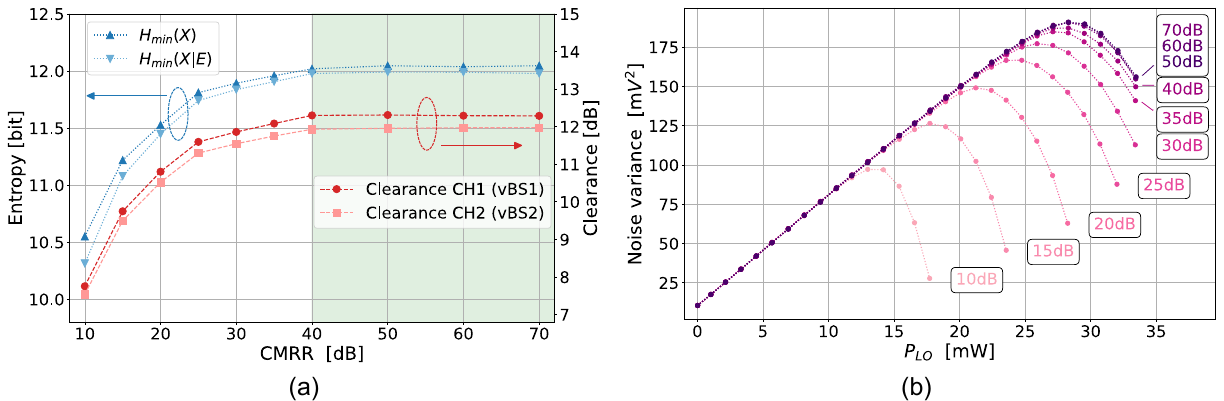}
    \caption{\textbf{\gls{qrng} performance and receiver characterization as a function of the receiver's \gls{cmrr}}. a) The blue curves represent the classical min-entropy $H_{min}(X)$ and the lower-bound on the quantum conditional min-entropy $H_\text{min}(X \vert E)$ for different \gls{cmrr} values. In red, the clearance of the two output channels of the heterodyne receiver are shown. Data were collected at the highest LO power compatible with detector linearity. The green region highlight CMRR values $\geq 40~\si{dB}$, which we identified as the operational range suitable for randomness generation.
    b) Receiver characterization obtained for different \gls{cmrr} values. The plot shows the characterization of output channel CH2, related to vBS2. Higher \gls{cmrr} values expand the exploitable linear region for randomness generation, allowing for higher clearance and rate. For CMRR values above $40~\si{dB}$, no further expansion of the \gls{bpd} linear region is observed. The noise variance data are extracted from heterodyne detections filtered in the $0.5-2.3~\si{\giga\hertz}$ frequency range, satisfying the flat-spectra requirement for secure randomness generation.}
    \label{fig: min_entropy_calib_line}
\end{figure*}

A schematic representation of the experimental setup used to implement the \gls{qrng} protocol is shown in Fig.~\ref{fig: optical setup}.
The laser light is divided between the Signal and \glsxtrlong{lo} line, where VOA-1 is only required for calibrating the receiver by estimating the $m_{x,p}$ coefficients.
Meanwhile, the Signal line is detached from the PIC input to guarantee the injection of the vacuum state into the \gls{qrng}.
Although it is crucial to know which state we are using during calibration, its nature is not essential when running the protocol, since it operates in the source-device independent framework.
Therefore, we opted for the vacuum state as the device input also for randomness generation, given the simplicity with which it can be realized.

At the receiver end, the optical signals extracted by the \gls{pic} are converted into electrical ones by a pair of \glspl{bpd}.
These signals are then further filtered by additional $\SI{50}{\mega\hertz}$ RF \glsxtrlongpl{hpf} to improve low-frequency noise rejection.
Finally, the resulting heterodyne measurements are digitized by an 8-bit oscilloscope with a frame duration of $400~\si{\micro\second}$, and subsequently transferred to a computer for off-line digital processing, accounting for \gls{dsp} and randomness extraction.
The former is based on~\cite{Bertapelle:25} and is used to prepare the data for either QRNG performance estimation or randomness distillation.
The \gls{dsp} extracts and isolates via digital re-sampling a portion of the measurement spectrum, which must be sufficiently flat to avoid temporal correlations in the random string.
According to such, we selected a spectral window range of $0.5-\SI{2.3}{\giga\hertz}$.
This choice also rejects the technical noise present at lower frequencies, is sufficiently far from the BPDs' upper cutoff frequency of \SI{2.5}{\giga\hertz}, and ensures that the signals possess a clearance of at least \SI{12}{\dB} (see Appendix \ref{App:B.2}).
To extract secure true randomness, the raw data, pre-processed with the methods outlined above, are hashed with a 2-universal hashing function \cite{Tomamichel2011}.
The latter is implemented as a binary matrix-vector multiplication, and the final data are analyzed with the standard NIST test suite.

\subsection{Results}
After the \gls{pic} was configured as a heterodyne receiver and characterized, we determined the performance that can be achieved by the \gls{qrng} and reported it in Fig.~\ref{fig: min_entropy_calib_line}(a).
We show the quantum conditional min-entropy $H_\text{min} (X \vert E)$, which quantifies the extractable truly random bits from a 16-bit heterodyne measurement (8-bit per channel), as a function of the receiver's \gls{cmrr}.
Notice that as the \gls{cmrr} increases, the value of $H_\text{min}(X \vert E)$ also increases.
However, for higher values, the improvements become progressively smaller, and a \gls{cmrr} greater than \SI{40}{\dB} yields no significant gain in $H_\text{min}(X \vert E)$.
Therefore, if we consider a \gls{cmrr} of at least \SI{40}{dB}, the \gls{qrng} presented here lower-bounds $H_\text{min}({X \vert E})$ to $\sim \SI{11.98}{\bit}$ (corresponding to a phase-space resolution of $\delta_x \sim 2.94\cdot 10^{-2}$ and $\delta_p \sim 2.65\cdot 10^{-2}$), thus achieving a secure generation rate of:
\begin{equation}
    R_{sc} = R_{\rm raw} H_\text{min} \left( X\vert E \right) \approx \SI{43.13}{Gbps}\, ,
\end{equation}
with $R_{\rm raw}=3.6~\si{\giga\sample\per\second}$ the equivalent ADC sampling-rate after digital re-sampling, which is performed off-line.
It is worth mentioning that the gain in terms of the quantum conditional min-entropy is due to the expansion of the \glspl{bpd} linear region due to the higher \gls{cmrr}.
This allows for the exploitation of a greater local oscillator power while the coefficient $m_{x,p}$ remains unchanged.
Fig.~\ref{fig: min_entropy_calib_line}(b) shows the behavior just outlined for one output channel, while the other acts similarly.
From the acquired data, we estimated a coefficient of $m_x=(5.68\pm 0.01)\cdot 10^{-3}~\si{\volt\squared\per\watt}$ for the photodetector attached to vBS1 and $m_p=(6.95\pm 0.02)\cdot 10^{-3}~\si{\volt\squared\per\watt}$ for the photodetector attached to vBS2, reported in Fig.~\ref{fig: min_entropy_calib_line}(b).
In Fig. \ref{fig: min_entropy_calib_line}(a), we also reported the device's classical min-entropy $H_\text{min}(X)$, a quantity that upper-bounds $H_\text{min}({X \vert E})$ and represents the number of truly random bits that can be extracted by considering a fully trusted randomness generator.
As can be seen, the gap between the latter quantities becomes smaller with increasing \gls{cmrr} values because the device's clearance improves.
This reduces the impact of classical noise, leading to better phase-space quadrature resolution and higher generation rate.
Notice that even at low \glspl{cmrr}, the gap is small, highlighting the capabilities of our heterodyne receiver.

Following the characterization of the \gls{qrng} receiver, we collected approximately \SI{500}{\giga\byte} of data for random number extraction.
For this purpose, we set $P_{\rm LO}$ to \SI{22.5}{\milli\watt} which corresponds to the maximum value allowing for linear operation of the \glspl{bpd}, given a \gls{cmrr} of at least \SI{40}{\dB}.
After the acquisition and pre-processing, which accounts for spectral isolation through digital re-sampling, we distilled random numbers with a 2-universal hashing function implemented as a Toeplitz bit-wise matrix-vector multiplication.
The latter used $n=28532$ rows and $m=38208$ columns resulting in an individual-run epsilon-security parameter of the order of $\varepsilon \sim 10^{-40}$ and an actual generation-rate:
\begin{equation}
    R_{sc} = \left( \dfrac{n}{m} \right) n_\text{bit} R_{\rm  raw} \approx 43.01~\si{Gbps}\, .
\end{equation}
where $n_\text{bit}=16$ is the number of bit acquired for each heterodyne outcome.
Hence, with such choices concerning pre-processing and the matrix dimensions, we obtained approximately \SI{48}{\giga\byte} of hashed data from the original $\sim\SI{500}{\giga\byte}$ of raw heterodyne measurements with an overall security parameter $\varepsilon^\prime \sim 10^{-33}$.
Subsequently, the processed random numbers were analyzed using NIST standard random test battery suits, with the results reported in Appendix \ref{App:NIST}.
The results did not reveal any significant non-ideality in the randomness of the generated numbers.

\section{CV-QKD Configuration}\label{Sec:QKD}
In this section, we evaluate the capabilities of our \gls{pic}, configured as a heterodyne receiver, for \gls{cv}-\gls{qkd} applications.
We opted for the discrete-modulated coherent states protocol~\cite{denys2021}, specifically employing the \glsxtrlong{qpsk} constellation.
Compared to the Gaussian modulation technique \cite{grosshans2002}, the constellation-based method mitigates some implementation limitations like the need for a continuous Gaussian state encoder, low reconciliation efficiency, and computationally demanding error correction procedures~\cite{almeida2021}, albeit at the cost of a reduced performance in terms of \gls{skr}.
The latter is computed by using the Devetak-Winter bound~\cite{devetak2005}, expressed as \mbox{$\text{SKR}=\beta \, \text{I}_{\text{AB}}-\chi_{\text{BE}}$}, where $\beta$ is the error correction efficiency, $\text{I}_{\text{AB}}$ is the mutual information between Alice and Bob, and $\chi_\text{E}$ is the Holevo bound, which upper-bounds the information accessible to Eve about Bob’s data in the reverse reconciliation scenario.
Notice that in our setup, Alice is represented by the Source and Signal Line without ATT and the Quantum Channel, Bob comprises the LO Line and Optical Rx.
Although ${\rm I}_{\text{AB}}$ can be determined from Alice's and Bob's data,
the Holevo bound is computed as the worst-case scenario compatible with Bob's measurement.
To this end, we used the theoretical framework of~\cite{denys2021} and extended it to account for the trusted detector assumption, where the efficiency $\eta$ and the electronic noise $V_{\text{el}}$ of the receiver are characterized and considered inaccessible to an eavesdropper (this work will be detailed in a future publication).
A detailed explanation of the receiver characterization procedure is given in Appendix~\ref{App:calibration}.
At each round of the protocol, the covariance matrix of Alice and Bob's data $\expval{\text{X}_\text{A} \text{X}_\text{B}}$ is computed and Alice's variance $V_A$ is estimated from the optical power $P_{\rm tx}$ measured by PM-2 as twice the mean number of photon $\langle n \rangle$ per symbol:
\begin{equation}
    V_A = 2\cdot\langle n \rangle = 2\, \frac{P_{\rm tx}}{E_{\rm ph} R}\, ,
\end{equation}
where $E_{\rm ph}$ is the photon's energy and $R$ is the system's repetition rate.
Then, $\chi_\text{E}$ is bounded by estimating the channel's parameters, namely the transmittance $T$ and the excess noise $\xi_A$ reported at Alice's side, and using the calibrated values $V_A, \eta$ and $V_{\rm el}$.
The channel transmittance $T$ and excess noise are obtained from the following relations:
\begin{align}
    & \expval{\text{X}_\text{A} \text{X}_\text{B}} = \sqrt{\frac{\eta T}{2}} V_A \label{eq:cvqkd-cov}\\
    &V_B = 1 + V_{\rm el} + \frac{\eta T}{2} V_A + \frac{\eta T}{2} \xi_A\label{eq:cvqkd-var}\, ,
\end{align}
while the receiver losses $\eta$ are derived from detector efficiency (Appendix~\ref{App:calibration}) and PIC insertion losses.

\subsection{Experimental setup}
The setup used to assess the \gls{pic}'s performance for \gls{cv}-\gls{qkd} is shown in Fig.~\ref{fig: optical setup}.
The Signal-line phase encoder generates a \SI{250}{\mega\baud} optical \gls{qpsk} signal.
The modulating signals provided by the \gls{fpga} are shaped with a \gls{rrc} filter with a roll-off factor of $0.3$ to ensure efficient use of the electronic bandwidth.
Then, the variable optical attenuator VOA-2 is used to optimize the modulation variance $V_A$, considering the additional attenuation provided by the fixed optical attenuator ATT.
Moreover, the \glsxtrlong{pc} PC-2 ensure that the polarization state of both signal and \gls{lo} are matched at the \gls{pic}'s input.
To estimate $V_A$, \SI{99}{\percent} of the optical power is tapped with BS-2 and measured by the \glsxtrlong{pm} PM-2.
Since both the signal and the \gls{lo} originate from the same laser, we used a fiber delay of length $\Delta L$ to match both optical paths, reducing phase instabilities.
The choice to rely on a shared laser simplifies the experimental implementation of the setup.
Although the \gls{llo} technique is preferred for actual \gls{cv}-\gls{qkd} implementations due to security, performance, and practical reasons~\cite{qi2015}, the shared approach is still adequate to evaluate the performance of \gls{cv}-\gls{qkd} systems.
However, we note that our heterodyne receiver can, in principle, be integrated with a \gls{cv}-\gls{qkd} system exploiting the \gls{llo} technique, as the additional complexity arises mainly at the \gls{dsp} stage and is independent of the specific receiver technology adopted.

Upon entering the heterodyne receiver through the input port, the quantum signal interferes with the \gls{lo}, and its optical quadratures are extracted by a pair of \glspl{bpd}.
Such electrical signals are then digitized by the oscilloscope and streamed to a computer for offline \gls{dsp}, data analysis and parameter estimation.
The former accounts for a \SI{50}{\mega\hertz} \glsxtrlong{hpf} and a \gls{rrc} filter matching the one used at the transmitter side to reject technical noise and to minimize inter-symbol interferences respectively.

\subsection{Results}
The main parameters of our \gls{cv}-\gls{qkd} system are summarized in Table~(\ref{tab:results-parameter-estimation}).
Since we assumed a trusted detector scenario, the overall system losses $\eta$ were $0.55$, and the electronic noise 0.029 SNU (Shot Noise Units), corresponding to a receiver clearance of approximately \SI{16}{dB}. 
The electronic noise variance was measured when the laser was turned off and the shot noise variance was obtained by injecting light only into the LO port while the signal input was attenuated by VOA-2 by $\sim 80~\si{dB}$.
More details can be found in Appendix~\ref{App:calibration}.
\begin{table}
    \centering
    \renewcommand{\arraystretch}{1.2} 
    \begin{tabular}{cccc}
        \hline \addlinespace
        \textbf{Parameter} & \textbf{Symbol} & \textbf{Value} & \textbf{Units} \\ \hline \addlinespace
        \makecell{Electronic noise \\ variance} & $V_{\rm el}$ & $0.029$ & \si{\snu} \\ \hline
        Receiver efficiency & $\eta$ & $0.55$ & \\ \hline
        Channel transmittance & $T$  & $0.73$ & \\ \hline
        Alice's variance & $V_{A}$ & $0.46$ & \si{\snu} \\ \hline
        Excess noise & $\xi_{A}$  & $0.015$ & \si{\snu}\\ \hline
        Asymptotic & \multirow{2}{*}{$\text{SKR}$}& $0.013$ & \si{\bit\per\symbol} \\
        Secret Key Rate & & $3.2$ & \si{\mega\bit\per\second} \\ \hline
    \end{tabular}
    \caption{\textbf{Summary of the parameters of the \gls{cv}-\gls{qkd} experiment.} Results of parameter estimation employing a \SI{250}{\mega\baud} \gls{qpsk} modulation and with reconciliation efficiency set to \SI{95}{\percent}.
    The results are obtained by analyzing 
    $50$ acquisitions, each with an acquisition time of $1~\si{\milli\second}$ and $250\,000$ recorded symbols.}
    \label{tab:results-parameter-estimation}
\end{table}
It is worth noting that one of the key advantages of \gls{flm} technology for \gls{cv}-\gls{qkd} is the low optical losses that such a platform can grant.
The modulation variance $V_A$ of the symbols transmitted by Alice was measured using the PM-2 power meter and taking into account the $\SI{30}{\dB}$ optical attenuator ATT.
The obtained value is \SI{0.46}{\snu}. 
Then, the excess noise parameter $\xi_A$ and channel transmittance $T$ were determined using Equations~(\ref{eq:cvqkd-cov}) and~(\ref{eq:cvqkd-var}) with $\expval{\text{X}_\text{A} \text{X}_\text{B}}$
and $V_B$ estimated by analyzing a dataset of $\num{12.5e6}$ symbols acquired with the oscilloscope.
The values obtained were respectively $\xi_A=0.015~\si{\snu}$ and $T=0.73$.
Finally, the asymptotic \gls{skr} reached by the system is approximately \SI{0.013}{\bit\per\symbol}, corresponding to $\sim\SI{3.2}{\mega\bit\per\second}$ given the \SI{250}{\mega\baud} symbol rate.

\begin{figure}
	\centering
	\includegraphics[width=1\linewidth]{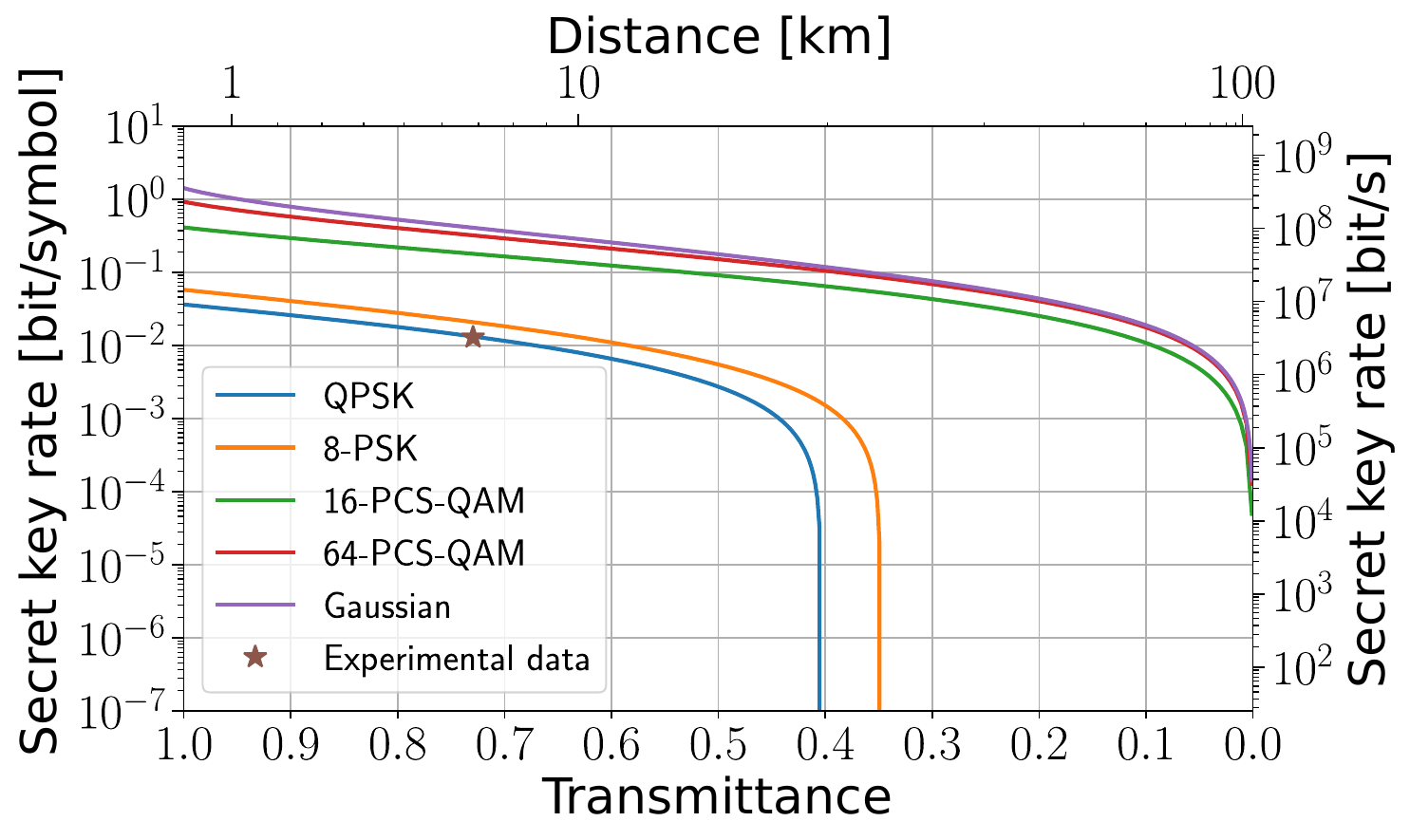}
	\caption{\textbf{Simulated \gls{cv}-\gls{qkd} performance with different modulation formats}. Simulation of the asymptotic secret key rate for different \gls{cv}-\gls{qkd} protocols, using the experimental parameters obtained from the parameter estimation reported in Table~\ref{tab:results-parameter-estimation}.
    For each transmittance value, Alice's variance was optimized to maximize the secret key rate.
    Additionally, for \gls{pcs}-\gls{qam} protocols, the probability distribution parameter was optimized as well.}
    \label{fig:skr-simulation}
\end{figure}

In our \gls{cv}-\gls{qkd} implementation, we adopted a \gls{qpsk} modulation scheme, which is relatively straightforward to implement.
However, this simplicity comes at the cost of a lower secret key rate compared to more sophisticated modulation schemes.
As a matter of fact, the best results are obtained using Gaussian modulation, and the latter performance can be approached with high-density discrete constellations.
Therefore, to better assess the potential of our system, we simulated the \gls{skr} that can be achieved as a function of transmittance, under the assumption that $\xi_A$, $V_{\rm el}$ and $\eta$ do not change and by adopting the optimal $V_{A}$ for each attenuation.
The results are shown in Fig.~\ref{fig:skr-simulation}, where we considered \gls{qpsk}, 8-\gls{psk}, 16-PCS-QAM, 64-PCS-QAM, and Gaussian as alternative approaches.
As can be seen, by increasing the constellation dimension of a \gls{psk}-based modulation format, performance can improve, although the gains are marginal compared to what can be obtained with the \gls{pcs} \gls{qam} or Gaussian modulation technique.
Implementing these higher-performance schemes would require incorporating an intensity modulator, which we plan to investigate in a future work to enhance our system’s capabilities.

\section{Conclusions}\label{Sec:Conclusion}
In this work, we demonstrated the successful implementation of a glass-based \gls{pic} specifically designed to serve as a versatile coherent receiver for CV quantum information protocols.
Manufactured using the \gls{flm} technique, the \gls{pic} offers several advantages, including a compact footprint, low overall losses, and insensitivity to input polarization. 
These characteristics enable seamless integration with \gls{cots} components, such as lasers and wide-band detectors, making it a cost-effective solution for \gls{cv} quantum information and communications, with promising strategies for integration of active elements, quantum light sources, and detectors already demonstrated \cite{spring2013chip, hou2021waveguide, gualandi2025}.
The device's high tunability, enabled by the integrated TOPSs, ensures adaptability to different operational conditions. After proper calibration, it supports the implementation of a heterodyne receiver with precise $\pi/2$ quadrature phase-shift and high \gls{cmrr} ($> \SI{73}{\dB}$).
These performance metrics enabled the optimal operation of the coherent receiver in both \gls{cv}-\gls{qrng} and \gls{cv}-\gls{qkd} applications.
In particular, our \gls{qrng} device achieved a secure generation rate of $\SI{42.74}{\giga\bit\per\second}$, which, to the best of our knowledge, represents a record-high result within the source-device-independent framework.
On the other hand, our \gls{cv}-\gls{qkd} system, employing a \gls{qpsk} modulation format, reached a SKR of \SI{3.2}{\mega\bit\per\second}.
Additionally, the \gls{pic} exhibited excellent operational stability over extended use, further supporting its suitability for practical quantum communication deployments.

Through this work, we benchmarked and demonstrated the suitability of \gls{flm}-fabricated glass \glspl{pic} for \gls{cv} quantum communication, demonstrating their effectiveness for coherent heterodyne detection.
Besides clear advantages in terms of compactness for integrated solutions, the use of glass offers an intrinsically inert material, highly resistant to environmental perturbations. Moreover, waveguides buried in 3D within the volume are naturally shielded, a feature that provides circuit resilience even under the harsh conditions of space missions \cite{Piacentini2021, ahmadi2024quick3}, further strengthening the real-world relevance of \gls{flm}-fabricated devices.

\acknowledgments

We acknowledge financial support from the European Union – NextGenerationEU, within the PNRR MUR project National Quantum Science and Technology Institute – NQSTI (PE0000023-Spoke 7).

This work was partially supported by European Union's Horizon Europe research and innovation program under the project Quantum Secure Networks Partnership (QSNP), grant agreement No 101114043. Views and opinions expressed are however those of the authors only and do not necessarily reflect those of the European Union or European Commission-EU. Neither the European Union nor the granting authority can be held responsible for them.


\section*{Authors contributions}
GG manufactured and calibrated the photonic device and AP performed the optical characterization.
TB supervised the optical characterization.
TB and AP performed the QRNG experiment.
DM and TB performed the NIST tests.
MS, AP and YP performed the QKD experiment. YP compared the results with those of other state-of-the-art PICs.
GC and RO supervised the chip design and fabrication process.
DM, GV, PV and MA supervised the quantum application experiments.
MA, GV, PV and RO conceived the initial concept and methodology.
AP and TB wrote the initial draft of the manuscript. 
All authors discussed the methods, the results and contributed to the writing of the final manuscript.
\bibliography{main_bib}
\newpage
\onecolumngrid
\appendix
\section{PIC configuration methods}\label{App:pic_config}
In this appendix, we describe the procedures used to characterize the heterodyne configuration of the PIC, specifically the balancing of the two output channels and the phase shift applied to the local oscillator to ensure a \SI{90}{\degree} hybrid operation.

\medskip

Concerning the first point, the architecture of the glass-based PIC provides the capability to tune the splitting ratio of the \glspl{vbs} through its integrated \glspl{tops}.
Considering a single balanced detector connected to a \gls{vbs}, let $i_1$ and $i_2$ denote the photocurrents generated by the \gls{bpd}'s photodiodes in the balanced configuration ($i_1\sim i_2$).
According to the standard definition of CMRR as the ratio between the differential and common-mode gain, such quantity is derived as:
\begin{equation}
    \text{CMRR} = 20\,\log_{10}\left(\frac{(i_1+i_2)/2}{\abs{i_1-i_2}}\right)
    =
    -20\,\log_{10}\left( 2\abs{\dfrac{i_1}{i_1+i_2}-\dfrac{i_2}{i_1+i_2}} \right) = -20\,\log_{10}\left(2\abs{2r-1}\right)    \label{eq:CMRR_def}
\end{equation}
where $r=\frac{i_1}{i_1+i_2}$ corresponds to the photocurrent splitting ratio.
To achieve optimal performance, a balanced receiver requires the maximization of the \gls{cmrr}, which occurs when $i_1\sim i_2$ (\textit{i.e.}, $r\sim 0.5$), corresponding to a 50:50 \gls{bs} under the condition of equal photodiode gains.
However, in practical implementations, slight mismatches in photodiode responsivities and variations in PIC-to-BPD coupling efficiencies may lead to uneven photocurrents, even with a perfectly balanced \gls{vbs}.
To address this, the PIC allows for the compensation of such non-idealities by acting on the integrated \gls{tops}.

\begin{figure}[t!]
    \centering
    \includegraphics[width=1.\linewidth]{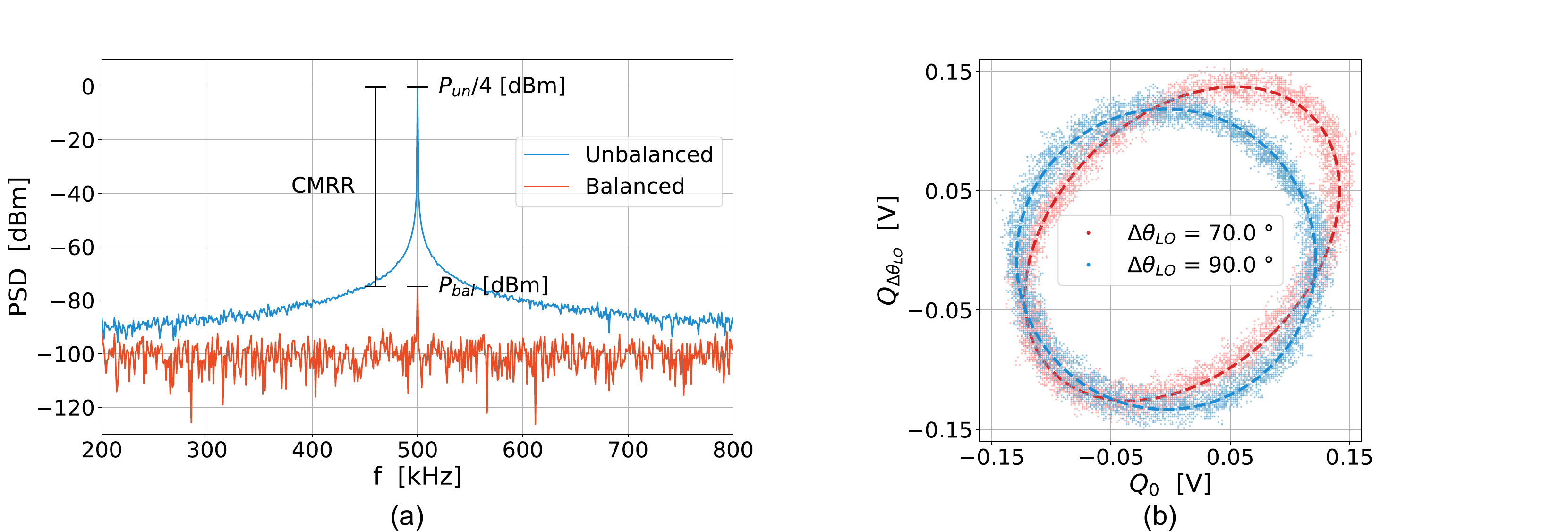}
    \caption{\textbf{Example of a PIC heterodyne configuration measurement}.
    (a) Power Spectral Density (PSD) of CH2 \gls{bpd} signals, connected to the output of vBS2, shown for the unbalanced (blue) and balanced (red) configuration.
    The \gls{cmrr} is extracted from the difference in spectral power at the modulation frequency, yielding a \gls{cmrr} of \SI{74.6}{dB}.
    (b) Phase-space representation of the heterodyne quadrature detections acquired with two different \gls{lo} phase-shifts $\Delta\theta_{\text{LO}}$.
    The measured quadrature samples are shown as scatter points, with the corresponding elliptical fit indicated by dashed lines. For optimal $\Delta\theta_{\text{LO}}= 90~\si{\degree}$, the ellipse approaches a circular shape, indicating the detection of two orthogonal quadratures.}
    \label{fig: cmrr_hybrid_estim}
\end{figure}

In practice, the \gls{cmrr} is measured based on the \gls{bpd} electrical output, which is proportional to the photocurrent difference $i_1-i_2$.
The measurement is conducted in an AC-setting by modulating the optical power of the laser with a $f_{\rm mod}=500~\si{\kilo\hertz}$ sinusoidal waveform.
The modulated light is coupled in the PIC, where each \gls{vbs} split the optical power between its output arms according to a controllable splitting ratio.
The \glspl{bpd} electrical signals exhibit a residual modulation whose amplitude depends on the photocurrent imbalance.
To estimate the \gls{cmrr}, we consider the two limiting cases of optimal balance and complete unbalance.
The first condition is obtained by minimizing the $f_{\rm mod}$ component in the \gls{psd} of the \gls{bpd} signal.
Its minimization, achieved by finely tuning the \gls{tops} bias voltage with sub-millivolt resolution, yields a minimum spectral power $P_{\rm bal}$.
The unbalanced condition is obtained by directing the entire light to a single \gls{vbs} output arm, resulting in a spectral power $P_{\rm un}$.
The latter power levels are related to the quantities in Eq.~(\ref{eq:CMRR_def}) as follows:
\begin{equation}
    \begin{cases}
        P_{\rm bal}= \frac{G^2}{R}\cdot\abs{i_1-i_2}^2 \\
        P_{\rm un}= \frac{G^2}{R}\cdot i_1^2 = \frac{G^2}{R}\cdot\left(i_1+i_2\right)^2\hspace{2mm} \vert_{i_2=0}
    \end{cases}
\end{equation}
where $G$ denotes the transimpedance gain and $R$ the load impedance.
The ratio between $P_{\rm un}$ and $P_{\rm bal}$ defines the system’s \gls{cmrr} as:
\begin{equation}
    \text{CMRR} = 10\,\log_{10}\left(\frac{P_{\rm un}/4}{P_{\rm bal}}\right)\, ,
    \label{eq: CMRR_estimation}
\end{equation}
Fig.~\ref{fig: cmrr_hybrid_estim}(a) shows a representative \gls{psd} of signals corresponding to the balanced and unbalanced configurations, yielding power levels of $P_{\rm un}/4=-0.3~\si{dBm}$ and $P_{\rm bal}=-74.9~\si{dBm}$.
From these values, we obtain a \gls{cmrr} estimation of \SI{74.6}{\dB}.
The system’s maximum attainable \gls{cmrr} is constrained by the noise floor $\sigma_0$ in the balanced configuration, measured at $-97.7~\si{dBm}$ (Fig. \ref{fig: cmrr_hybrid_estim}(a)).
Since $P_{\rm bal}$ cannot fall below this limit, the maximum \gls{cmrr} and the corresponding parameter $\delta$ from Eq. \ref{eq: CMRR_model} (discussed in the main text) are given by:
\begin{align}
    &\text{CMRR}_\text{max}=P_{\rm un}/4\text{[dBm]}-\sigma_0=97.4~\si{dB} \\
    &\delta=10^{-\text{CMRR}_\text{max}/20}=1.3\cdot 10^{-5}
\end{align}

\medskip

The second characterization procedure involves estimating the \gls{lo} phase shift introduced by \gls{tops} R3, which is required to enable the detection of two orthogonal quadratures of the quantum state at the \gls{pic} input.
To this end, we employed a classical signal with a triangular phase modulation $\phi_t$ imposed by the phase encoder.
Assuming the \gls{lo} to be in-phase with the modulated signal in vBS1, the two \glspl{vbs} sample two different quadratures given by:
\begin{equation}
    \begin{cases}
        x:=Q_0&=V_1\cos\left(\phi_t\right) \hspace{2.2cm}\text{from CH1}\\
        y:=Q_{\Delta\theta_{\rm LO}}&=V_2\cos(\phi_t+\Delta\theta_{\rm LO})\hspace{1.02cm}\text{from CH2}
    \end{cases}
\end{equation}
As can be seen, in the desired case of $\Delta\theta_{\rm LO}=\pi/2$, the detected quadratures are orthogonal to each other.
The pair $(x, y)$ parametrizes an ellipse in the optical phase space, with the signal phase $\phi_t$ acting as the parametric variable.
The resulting ellipse satisfies the implicit equation \mbox{$Ax^2+By^2+Cxy+Dx+Ey=1$}, where the quadratic and cross terms are:
\begin{equation}
    \begin{cases}
        A&=(V_1\sin(\Delta\theta_{\rm LO}))^{-2} \\
        B&=(V_2\sin(\Delta\theta_{\rm LO}))^{-2} \\
        C&=-2(V_1V_2\sin(\Delta\theta_{\rm LO})\tan(\Delta\theta_{\rm LO}))^{-1}
    \end{cases}
    \label{eq: ellipse_params}
\end{equation}
From these coefficients, the \gls{lo} phase shift is computed as:
\begin{equation}
    \Delta\theta_{\rm LO} = \arccos\left(-\frac{C}{2\sqrt{AB}}\right)\, .
    \label{eq: theta_LO_estimation}
\end{equation}
Experimentally, $\Delta\theta_{\rm LO}$ is estimated by acquiring the voltage traces from the two \glspl{bpd}, which are then mapped in the phase space and fitted to an ellipse.
The fitted curve parameters are then used in Eq.~(\ref{eq: theta_LO_estimation}) to compute the phase shift.
With this method, we perform 20 subsequent acquisitions to obtain an average estimation of $\Delta\theta_{\rm LO}$, resulting in a measurement uncertainty of approximately $0.2~\si{\degree}$.

\section{Calibration\label{App:calibration}}
In this appendix, we provide the details concerning the characterization of the heterodyne receiver used for both \gls{qrng} and \gls{cv}-\gls{qkd} applications.
The analysis accounts for assessing the photodetector responsivity, and both electronic noise and clearance estimation.

\subsection{Photo-detectors efficiency}
The efficiency of the balanced photodetectors (Thorlabs PDB780CAC) was determined by a responsivity measurement.
We injected light with known power $P_{\rm in}$ in each photodiode  of the two \gls{bpd} separately, and the corresponding amplified output voltage $V_{\rm out}$ was recorded.
The responsivity was then estimated as: 
\begin{equation}
    \mathcal{R} = \frac{I_{\rm out}}{P_{\rm in}} = \frac{V_{\rm out}}{G_{\rm TIA}\cdot P_{\rm in}}\, ,
\end{equation}
with $I_{\rm out}=\frac{V_{\rm out}}{G_{\rm TIA}}$ the photocurrent generated by the photodiode, and $G_{\rm TIA}$ the \gls{tia} gain.
The quantum efficiency $\eta_{\rm {PD}}$ is then computed as:
\begin{equation}
    \eta_{\rm PD} = \frac{h\nu}{e}\mathcal{R}\, .
\end{equation}
where $h$ is the Planck's constant, $\nu$ is the optical frequency and $e$ is the elementary charge.
The measured efficiencies were \SI{85.12}{\percent}, \SI{71.68}{\percent}, \SI{78.12}{\percent} and \SI{71.10}{\percent} averaging at \SI{76.505}{\percent}, compatible with the typical efficiency of these detectors.

\subsection{Electronic noise and clearance}\label{App:B.2}
\begin{figure}
	\centering
	\includegraphics[width=.5\linewidth]{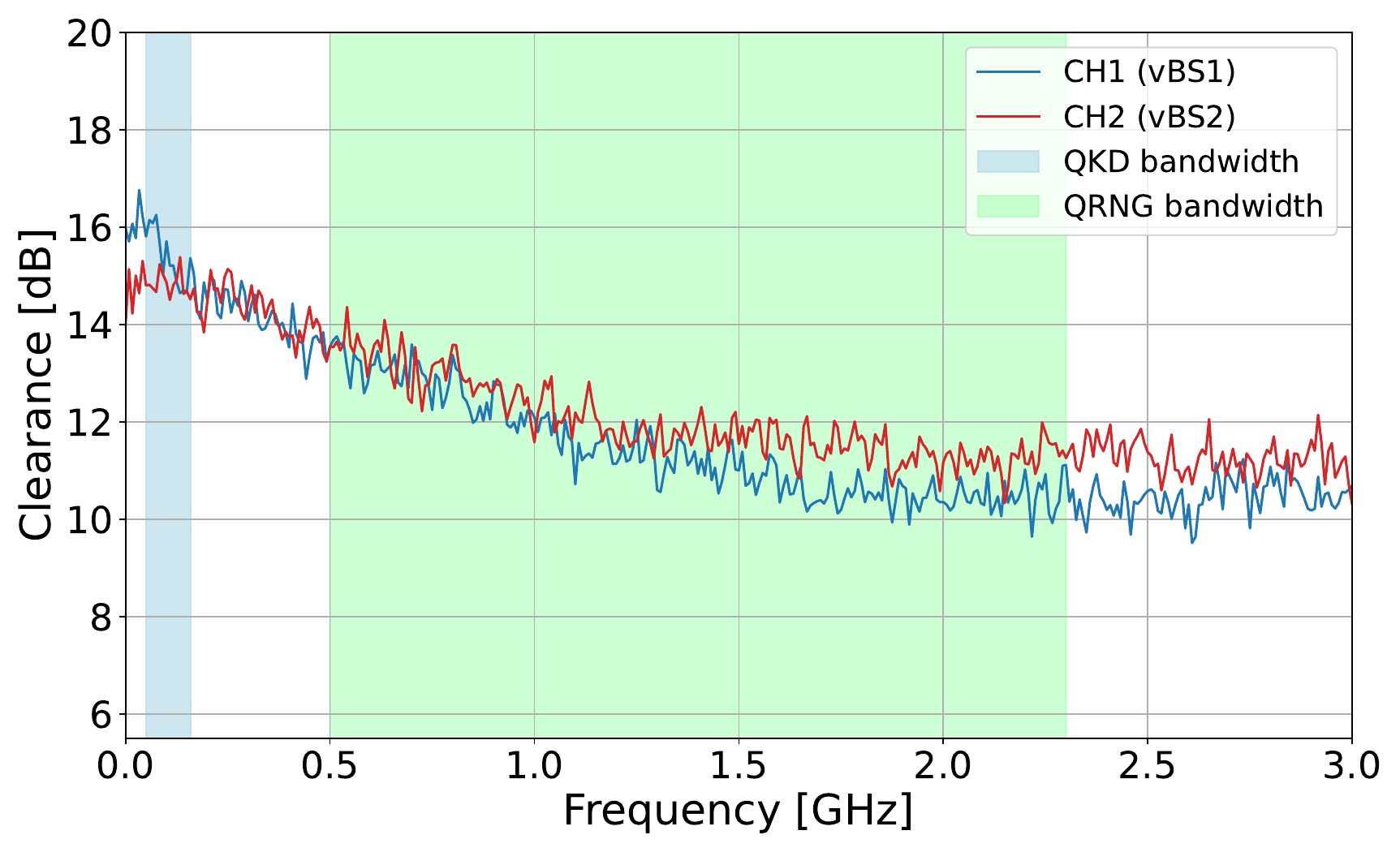}
	\caption{\textbf{Clearance of the receiver output channels}.
    The traces are obtained by subtracting the PSD of signals acquired with maximum LO power (compatible with detector saturation) and no LO power.
    The integral over the spectrum used for \gls{qrng} ($0.5-2.3~\si{\giga\hertz}$) and \gls{cv}-\gls{qkd} ($50-160~\si{\mega\hertz}$) applications provides the system overall clearance which is $\sim$ \SI{12}{dB} for the former case, and $\sim$ \SI{15}{dB} for the second.}
    \label{fig:characterization_and_clearance}
\end{figure}

The electronic noise of the receiver is estimated when no light is present at the device inputs (neither \glsxtrlong{lo} nor signal).
The receiver shot noise is acquired when only the \glsxtrlong{lo} is present.
By computing the difference in the \glsxtrlong{psd} associated to shot- (${\rm PSD}_{\rm LO}$) and electronic (${\rm PSD}_{\rm el}$) noise, we obtain the receiver clearance (Fig.~\ref{fig:characterization_and_clearance}):
\begin{equation}
    {\rm clearance} = 10\, \log_{10} \dfrac{{\rm PSD}_{\rm LO}}{{\rm PSD}_{\rm el}}
\end{equation}
As can be seen, this quantity varies across the bandwidth ranges chosen for \gls{qrng} ($0.5-\SI{2.3}{\giga\hertz}$) and \gls{cv}-\gls{qkd} (50–162.5~\si{\mega\hertz}) applications.

For \gls{cv}-\gls{qkd}, we ensured that the same \gls{dsp} used for the signal is also used for both the electronic and shot noise sequences to ensure the selection of the same mode.
The variance of the heterodyne measurements with and without \gls{lo}, $\sigma^2$ and $\sigma_{\rm el}^2$, are then computed, leading to an estimate of the shot noise variance $\sigma_0^2 = \sigma^2-\sigma_{\rm el}^2$.
Finally, the electronic noise $V_{\rm el}$ is determined as $V_{\rm el} = \sigma_{\rm el}^2/\sigma_0^2$ and the recovered data is normalized in shot noise units by dividing the data by $\sigma_0$.

\section{Randomness statistical tests}\label{App:NIST}
\begin{figure*}[h!]
	\centering
	\includegraphics[width=0.9\linewidth]{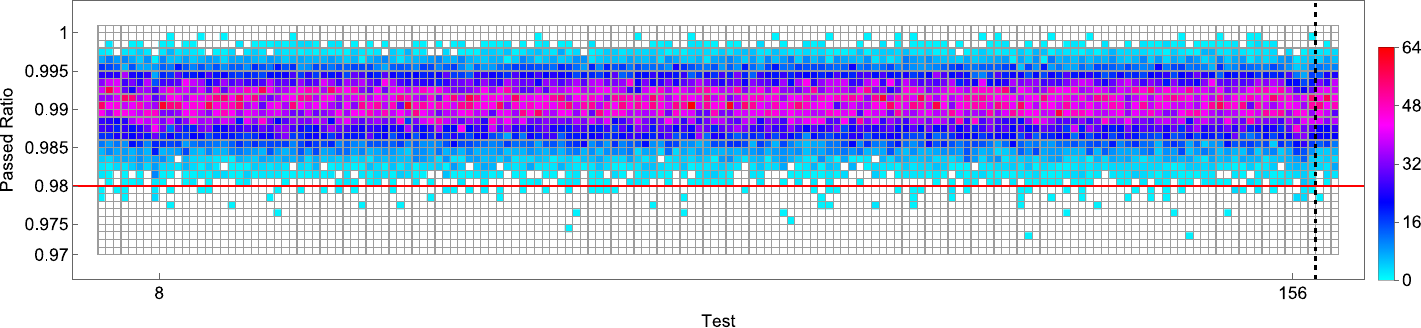}
	\caption{\textbf{Heatmap of the passing ratio for NIST battery applied to 345 files with a size of $\mathbf{10^9}$ bits.} 
    Columns correspond to the tests from 1 to 159 and from 186 to 188 (after the dashed lined). 
    Columns from 9 to 156 correspond to the \textsl{Non Overlapping Template} test. 
    The solid red line is the passing threshold. Tests from 160 to 185 are not reported because they feature a different threshold for each of the files.} 
	\label{fig:nist_passing_ratio}
\end{figure*}

\begin{figure*}
	\centering
    \includegraphics[width=0.8\linewidth]{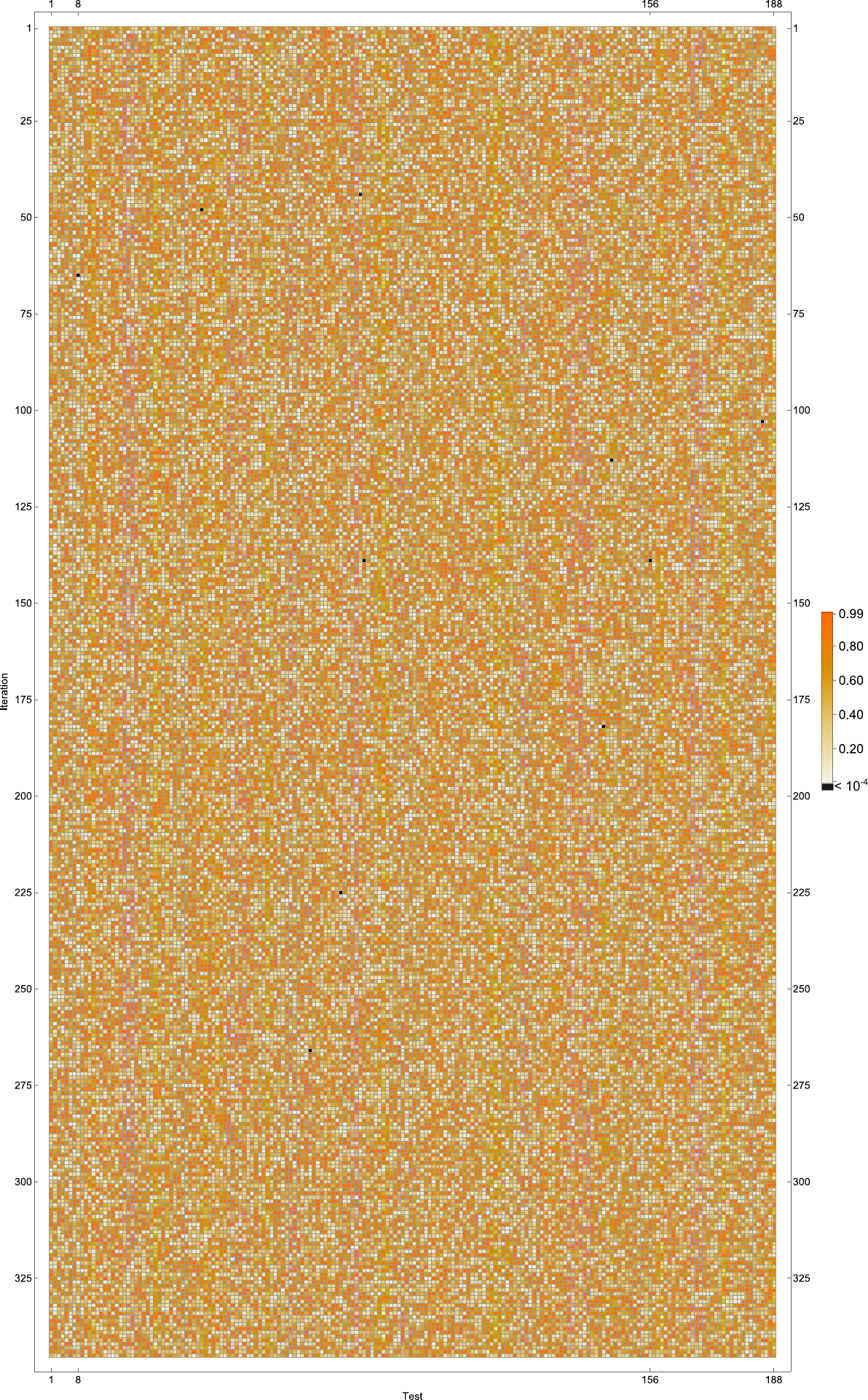}
	\caption{\textbf{P-values of the second order test on the uniformity of the p-values.} Each rows corresponds to a file analyzed, whereas each column corresponds to a test. A p-value lower than $10^{-4}$ is considered critical and is flagged in black.}
	\label{fig:nist_uniformity_all}
\end{figure*}

The random string generated by the presented QRNG and processed through a randomness extraction procedure was tested with the \textsl{NIST SP 800-22} battery of statistical tests for randomness assessment.
We generated 345 binary files with a length of $10^9$ bits.
The suite divides each file into $10^3$ substrings with a length of $10^6$ bits.
A total of 188 tests are then applied to the set of substrings and each test outputs a p-value.
For each test, two criteria are then used to assess whether a file has passed it. 
One checks that the proportion of substrings with a p-value larger than $10^{-2}$ is above a given threshold.
The other is a second order test checking whether the previous $10^3$ p-values are uniformly distributed: a chi-square test is applied to their distribution and the test is passed if the resulting p-value is larger than $10^{-4}$.
In Fig.~\ref{fig:nist_passing_ratio} we report a heatmap of the passing ratios for the files tested.
From the map we excluded tests from 160 to 185 corresponding to the \textsl{Random Excursion} and \textsl{Random Excursion Variant} test respectively, because they feature passing thresholds that vary from file to file. 
In Fig.~\ref{fig:nist_uniformity_all} we instead report the second-order uniformity p-values.
The results of the analysis did not indicate any suspicious behavior, as no tests consistently failed.

\end{document}